\documentclass[twoside]{cernyrep}
\usepackage{esint}
\usepackage[T1]{fontenc}
\usepackage[bookmarks, colorlinks=true, linktoc=page, pdftex, linkcolor=black, citecolor=black, urlcolor=blue]{hyperref}

\usepackage{fancyhdr}
\usepackage{subfig}
\usepackage[labelfont=bf]{caption}
\DeclareCaptionFont{tenpt}{\fontsize{10}{12}\selectfont}
\usepackage{url}

\fancyhfoffset{4 mm}
\fancypagestyle{ARTTITLE}{%
\fancyhf{} 
\lhead{\hfill CERN Accelerator School Proceedings ---
{\it RF for Accelerators} ---  Berlin,  Germany, 2023\hfill}
\lfoot{\hspace{3mm} Available online at \url{https://cas.web.cern.ch/previous-schools}}
\rfoot{\thepage\hspace*{3mm}}
 
}

\usepackage{varwidth}
\usepackage{xcolor}
\begin{document}
\title{Low Beta, Normal Conducting Cavities}
 
\author{Lars~Groening}

\institute{GSI Helmholtzzentrum f\"ur Schwerionenforschung, Darmstadt, Germany}


\begin{abstract}
The contribution is on issues being especially related to normal conducting cavities operating at non-relativistic beam energies. Various types of cavities are introduced w.r.t. their operation mode, application, advantages, and disadvantages. Special emphasis is put on their production and the challenges along the way from finalized Rf-design up to the operating cavity. This covers the choice of material, production, tolerances, alignment, cooling, and the demanding task of copper plating. The contribution closes with some remarks on Rf-commissioning and -conditioning.
\end{abstract}

\maketitle
\thispagestyle{ARTTITLE}
\section{Introduction}
These proceedings assume knowledge on basic features of Rf-cavities as properties of Maxwell's equations and equivalent resonant circuits. They focuse on issues specific to cavities designed for low energy beams and to be normal conducting. 

Low energy or low beta implies that the relative particle velocity $\beta = v/c$ does significantly depend on the particle's kinetic energy $E_k=m_0c^2(\gamma -1)$, where $\gamma = (1-\beta ^2)^{-1/2}$ and $m_0$ is the~particle's rest mass. This dependency weakens with increasing kinetic energy. Accordingly, the range of low beta is not defined strictly. However, this range is commonly considered to end at about~$\beta\approx $~0.3. One single cavity is operated at one unique frequency. The above dependence causes the cavity's geometry around the beam axis, namely the drift tubes and their distances from each other, to depend on the local design velocity $\beta $ of the beam. Hence, the cavity has a design being specific to the particle beam energy at the~cavity's unique position along the linac. Unlike high energy cavities, low energy beam cavities cannot be exchanged between different positions along the linac. This makes the provision of spare cavities practically impossible.

Low beta together with significant beam current raises the issue of considerable space charge forces from the repelling force between individual beam particles. This force scales with $(\beta\gamma)^{-3}$ and is defocusing in all spatial directions. To compensate space charge, the cavity design shall eventually provide for enhanced external focusing. This is accomplished by choosing a lower Rf-phase w.r.t.~crest and by providing more space for installation of transversely focusing quadrupoles. Additional longitudinal and transverse focusing has impact on the cavity geometry.

Normal conductivity causes Ohmic power loss at the inner cavity surface through surface currents. This loss comprises a considerable fraction of the incoming Rf-power ($n\cdot$10\%) and it is not available for beam acceleration. It increases the total wall-plug power and additionally, it must be cooled away properly in order to prevent the cavity from being de-tuned by thermal deformation or even from severe damage from local melting.

The finite surface resistance lowers the cavity's shunt impedance by a factor of about 10$^{\text{4}}$ compared to super-conducting cavities. Accordingly, the intrinsic resonance width is increased by the same factor. This leads to shortened time for the rise/fall of Rf-power, once the Rf-power has been turned on/off. These shortened times allow for operation of the cavities in pulsed mode. Rf-power levels may be changed within some ten milliseconds between consecutive Rf-pulses. Accordingly, the linac can switch between different ion species (mass-to-charge ratio) and serve practically as many-ion-species accelerator. Finally, normal conductivity does not require expensive and resource-demanding operation of infrastructure for provision of liquid nitrogen and helium.

These proceedings commence through introduction of a selection of cavity types followed by describing a process to determine what cavity type is appropriate for the specific project being considered. The fourth section is on various issues related to cavity production. Copper plating is a demanding part within cavity realization and is treated in a dedicated section. The contribution closes with summarizing the procedure of Rf-conditioning.

\section{Types of cavities}
This section gives a selection of the most commonly used types of normal conducting cavities which are used for acceleration or forming of longitudinal bunch properties of low energy ions. It is far from completeness and reflects the personal choice of the author.

\subsection{Pillbox}
\begin{figure}
	\centering\includegraphics[width=.65\linewidth]{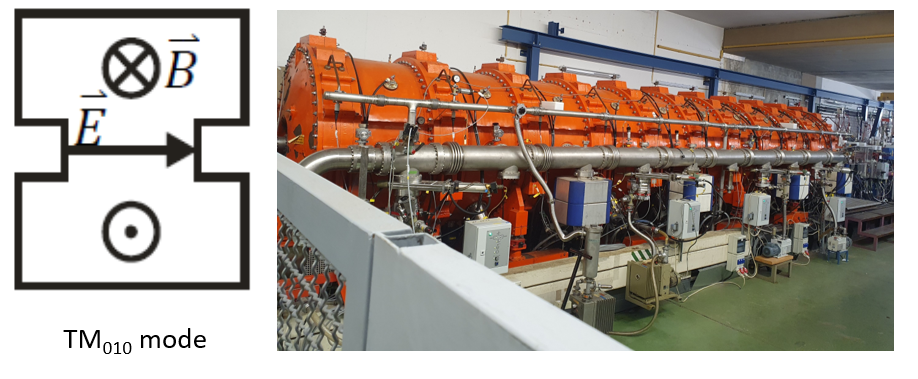}
	\caption{Left: scheme of electric and magnetic field inside of a pillbox cavity; right: ten single pill box cavities at GSI UNILAC operated at 108~MHz with gap voltage of up to~1.0~MV.}
	\label{fig:pillbox}
\end{figure}

The most simple type of cavity is the pillbox depicted in Fig.~\ref{fig:pillbox}. It resonates in the TM$_{\text{010}}$-mode and has just one single gap. Its shape is the most straightforward analogy to the resonant circuit. The capacitance is formed by the end plates and the two half drift tubes, being the start and end of the longitudinal electrical field. Surface currents between them flow along the cavity mantle and their path provides for the inductance. Inside of the two half drift tubes, elements for beam focusing or for diagnostics can be placed. Pillboxes mainly serve to change the shape of the longitudinal phase space ellipse, rather than for acceleration. Accordingly, re- and de-bunchers are often made from pillboxes. Sometimes they are applied for energy fine tuning. For instance, at GSI, a chain of 10 independent pillboxes provides for fine tuning of the energy to relative precision of 10$^{\text{-3}}$ for experiments at the nuclei's Coulomb barrier.

\subsection{\texorpdfstring{$\lambda$}{lambda}-Resonators}
\begin{figure}
	\centering\includegraphics[width=.75\linewidth]{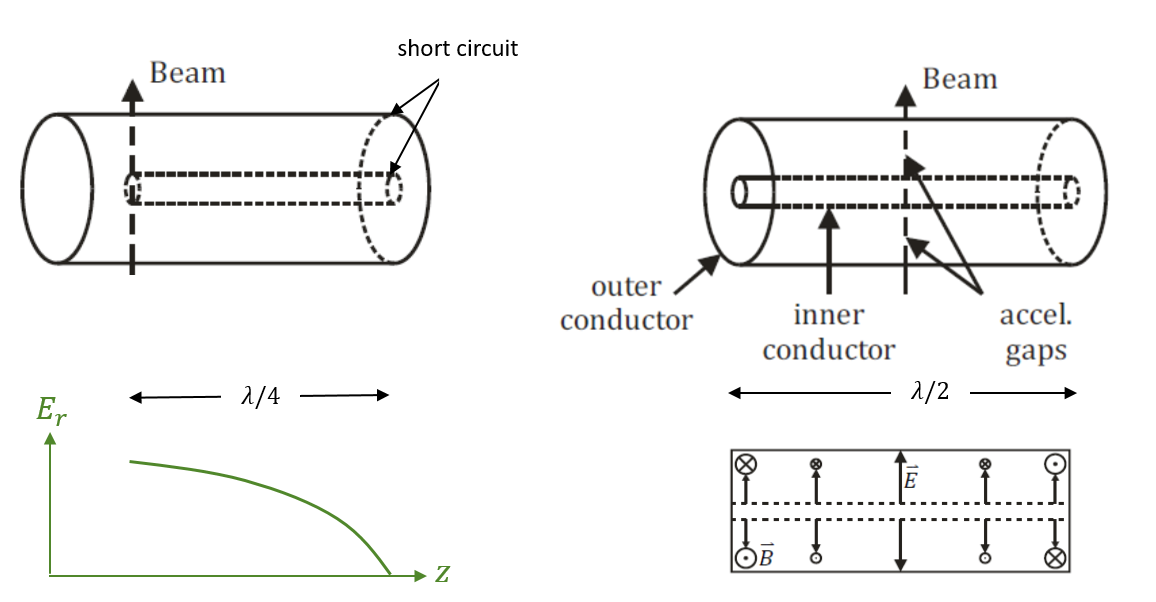}
	\caption{Scheme of \texorpdfstring{$\lambda$}{lambda}-resonators. Left:$\lambda/4$-resonator; right: $\lambda/2$-resonator.}
	\label{fig:lambda}
\end{figure}

Conceptually, the voltage between the inner and outer conductor of a coax cable can be used for acceleration or beam forming. In order to limit the resonator size and to provide for a standing Rf-wave, the two conductors need to be short circuited at least at one end. This short imposes boundaries to the electric field, i.e., at the short the electric field vanishes. Figure~\ref{fig:lambda} depicts schematically such a coax configuration.
The field along the coax axis is cosine-like and at a distance of $\lambda $/4 from the short, the field strength is at maximum. At this location, the two gaps are placed. In case of one single short, the resonator houses 1/4 of the wavelength and is called a $\lambda $/4-resonator, accordingly. The boundaries of of the electric field are
\begin{equation}
	E_r(z=L)\,=\,0\,\,\,\rightarrow\,\,\,L\,=\,(2n-1)\frac{\lambda}{4}
\end{equation}
and the resonant frequency is
\begin{equation}
	f\,=\,75\,\text{MHz}\cdot\frac{2n-1}{\text{L[m]}}\,.
\end{equation}
The frequency is doubled within the same resonator length if both ends are short circuited. For such a~$\lambda $/2-resonator the boundaries are
\begin{equation}
	E_r(z=0)\,=\,E_r(z=L)\,=\,0\,\,\,\rightarrow\,\,\,L\,=\,n\frac{\lambda}{2}
\end{equation}
with the frequency of
\begin{equation}
	f\,=\,150\,\text{MHz}\cdot\frac{n}{\text{L[m]}}\,.
\end{equation}
Single $\lambda $-resonators are used as bunchers at low energies. Being arranged in Rf-phase-locked chains, they comprise the first stage of ion linacs for inter-mediate and heavy masses. However, such chains are used generally in super-conducting linacs.

\subsection{Alvarez}
\begin{figure}
	\centering\includegraphics[width=.65\linewidth]{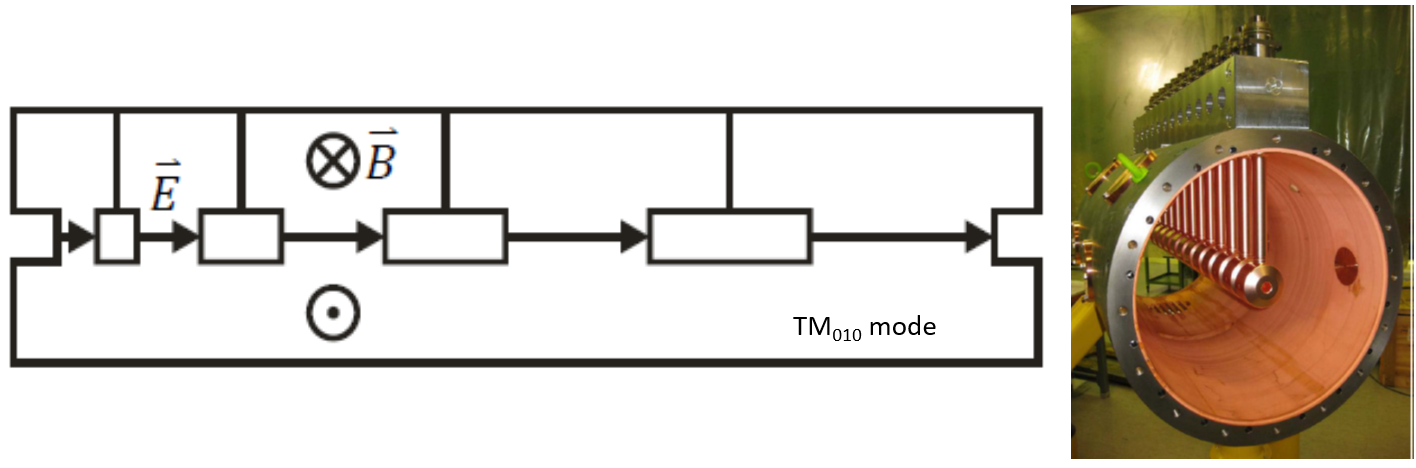}
	\caption{Left: scheme of electric and magnetic field inside of an Alvarez-type cavity; right: first DTL cavity of LINAC-4 at CERN operated at~352~MHz.}
	\label{fig:Alvarez}
\end{figure}

The Alvarez-type is the earliest type of cavity resonator comprising many gaps in order to provide for substantial acceleration. It resonates like a pillbox just that many drift tubes are integrated between the~half drift tubes of the pillbox. Although many drift tubes are integrated into the cavity, they lower the~frequency w.r.t. the empty cavity by just $\leq$10\%. The low change results from the fact that the~tubes comprise a chain of capacitances and accordingly, the chain's overall additional capacitance is low. The~distance between two consecutive gaps is given by~$\beta\lambda $, thus assuring a constant Rf-phase. An~Alvarez-cavity is one type of drift tube linac (DTL). Figure~\ref{fig:Alvarez} shows the field configuration and depicts the first Alvarez-cavity of CERN's LINAC-4 for protons~\cite{TDR_Linac4}. These cavities provide for very regular periodic focusing in the longitudinal and transverse plane. Usually, the Rf-phase is kept constant along all gaps; just at the beginning and end, small variations might be imposed in order to (re)-match the~beam (from) to the adjacent inter-cavity section. Additionally, the drift tubes are filled with one single quadrupole each, resulting into regular transverse beam envelopes. For this features, Alvarez-type cavities are generally chosen to provide beams being subject to substantial self-forces from space charge, i.e., high particle current at low energy.

\subsection{Spiral resonators}
\begin{figure}
	\centering\includegraphics[width=.45\linewidth]{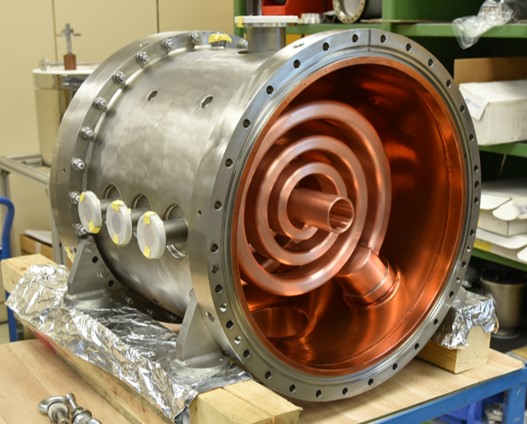}
	\caption{108~MHz spiral resonator of GSI's UNILAC.}
	\label{fig:spiral}
\end{figure}

The size of simple pillbox cavities increases inversely to the operation frequency, such that at low frequencies their diameter becomes too huge to be handled reasonably. In order to keep the cavity radius small at low frequency, the few drift tubes are suspended through a spiral~(Fig.~\ref{fig:spiral}). Doing so, the spiral augments the current path lengths and hence the overall inductance. 
Considering that the pill box cavity capacity scales with the square of its radius, the resonant frequency scales roughly as
\begin{equation}
	f\,\sim\,\frac{1}{\sqrt{LC}}\,\sim\,\frac{1}{\sqrt{L}R}\,.
\end{equation}
Accordingly, the frequency increase from the small size~$R$ is compensated by the enhancement of the~inductance~$L$ through the spiral suspension. Spiral resonators are used for bunching and for modest energy variation.

\subsection{H-mode}
\begin{figure}
	\centering\includegraphics[width=.65\linewidth]{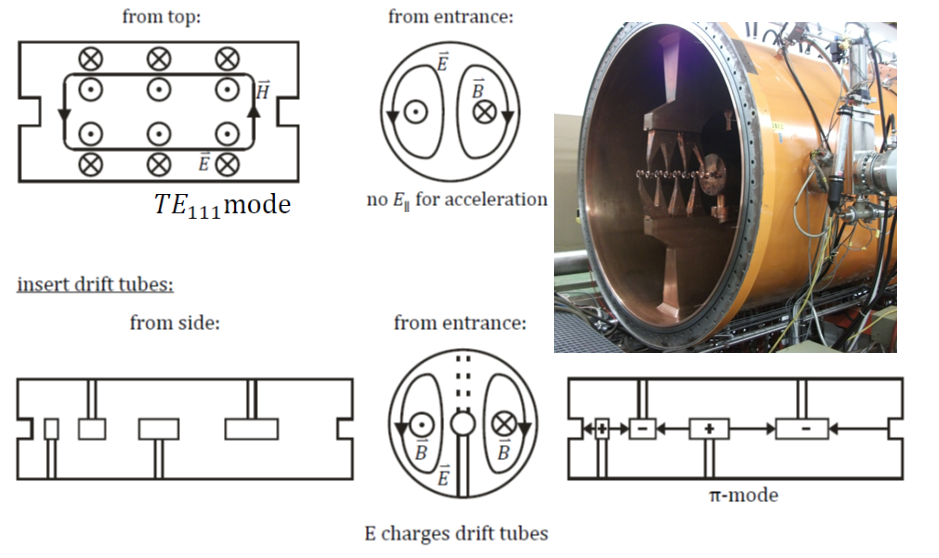}
	\caption{Scheme of electric and magnetic field inside of an IH-type cavity; upper right: 36~MHz inter-digital H-mode cavity of GSI's UNILAC.}
	\label{fig:IH}
\end{figure}

\begin{figure}
	\centering\includegraphics[width=.65\linewidth]{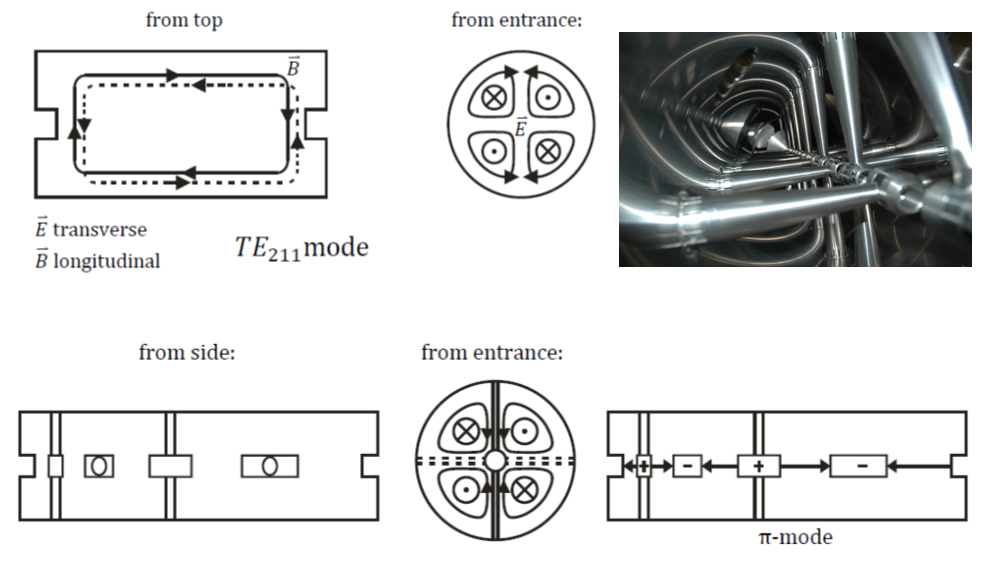}
	\caption{Scheme of electric and magnetic field inside of an CH-type cavity; upper right: 325~MHz crossed-bar H-mode cavity of the upcoming proton Linac at FAIR.}
	\label{fig:CH}
\end{figure}

Another type of DTL is made from cavities resonating in the H-mode (TE$_\text{n11}$) as shown at the left sides of Fig.~\ref{fig:IH} and Fig.~\ref{fig:CH}. A longitudinal magnetic field is formed along the initially empty cavity. The field lines are surrounded by the electric field lines.
Along the latter, stems are placed with alternating transverse orientation. They are closed by small drift tubes along the beam axis and by the electric field these are charged with alternating polarity along the beam axis. Accordingly, the distance between adjacent gaps at constant Rf-phase is reduced to~$\beta\lambda $2. Thanks to the small size of the drift tubes, the capacity formed by one gap is very small. Unlike as for an Alvarez-cavity, it is the transverse electric field along the cavity which is used to charge the drift tubes. This provides for enhanced gap voltages of H-mode cavities. Small capacitances together with high voltages strongly augment the shunt impedance of H-mode cavities compared to pillboxes and especially to Alvarez-DTLs. In turn, the small drift tubes do not allow for housing of focusing or diagnostic elements.

H-mode cavities resonating in TE$_\text{111}$-mode (Fig.~\ref{fig:IH}) are called inter-digital H-mode (IH) cavities~\cite{ratzinger_1996}, while the next higher magnetic mode (TE$_\text{211}$) is used in crossed-bar H-mode cavities~(CH) as shown in~Fig.~\ref{fig:CH}. IH-cavities are used for fast and efficient acceleration of ions with inter-mediate to high masses. CH-cavities enlarge the application of the latter to lightest ions at higher frequencies~\cite{Clemente}.

\begin{figure}
	\centering\includegraphics[width=.65\linewidth]{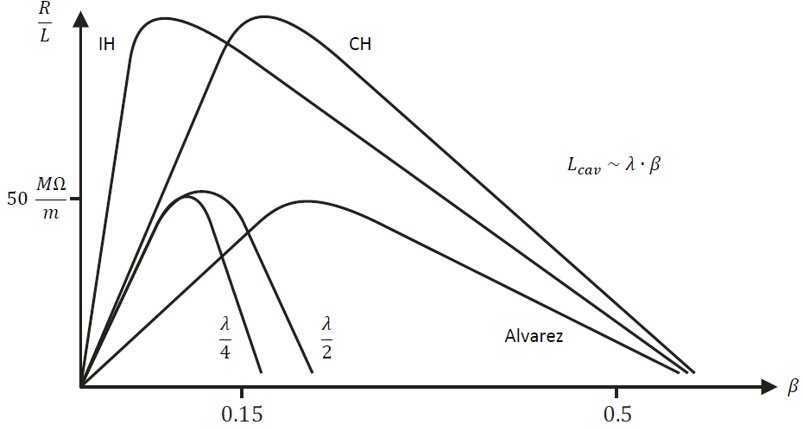}
	\caption{General scaling of the shunt impedance per length as a function of the relative beam velocity for various types of cavities.}
	\label{fig:merits}
\end{figure}

This section shall close with the qualitative comparison of shunt impedances from different cavity types. Figure~\ref{fig:merits} plots the shunt impedance per cavity length as a function of the relative particle design velocity at a given frequency. All curves have the same shape qualitatively but the positions and heights of the maxima differ according to the cavity type. H-Cavities have largest shunt impedances at lowest energies, since the drift tubes are charged by the transverse electric field. As the field can be "tapped" many times, this effect is most efficient if the drift tubes are close to each other, i.e., at low~$\beta$.

In general, very low $\beta$ implies short gap distances and hence large gap capacitances. These require more charge to be conducted onto the drift tubes to obtain a given voltage. In turn, these enlarged currents cause increased Ohmic losses. Instead, at very large~$\beta$, the gap capacity is small but the current path length between adjacent drift tubes becomes long. This also causes increase of the overall Ohmic losses. In consequence, there is a~$\beta _{max}$ with maximum shunt impedance in between. The curves plotted in Fig.~\ref{fig:merits} shift towards larger~$\beta$ if the frequency is augmented.

\subsection{Radio-frequency-quadrupole (RFQ)}
\begin{figure}
	\centering\includegraphics[width=.65\linewidth]{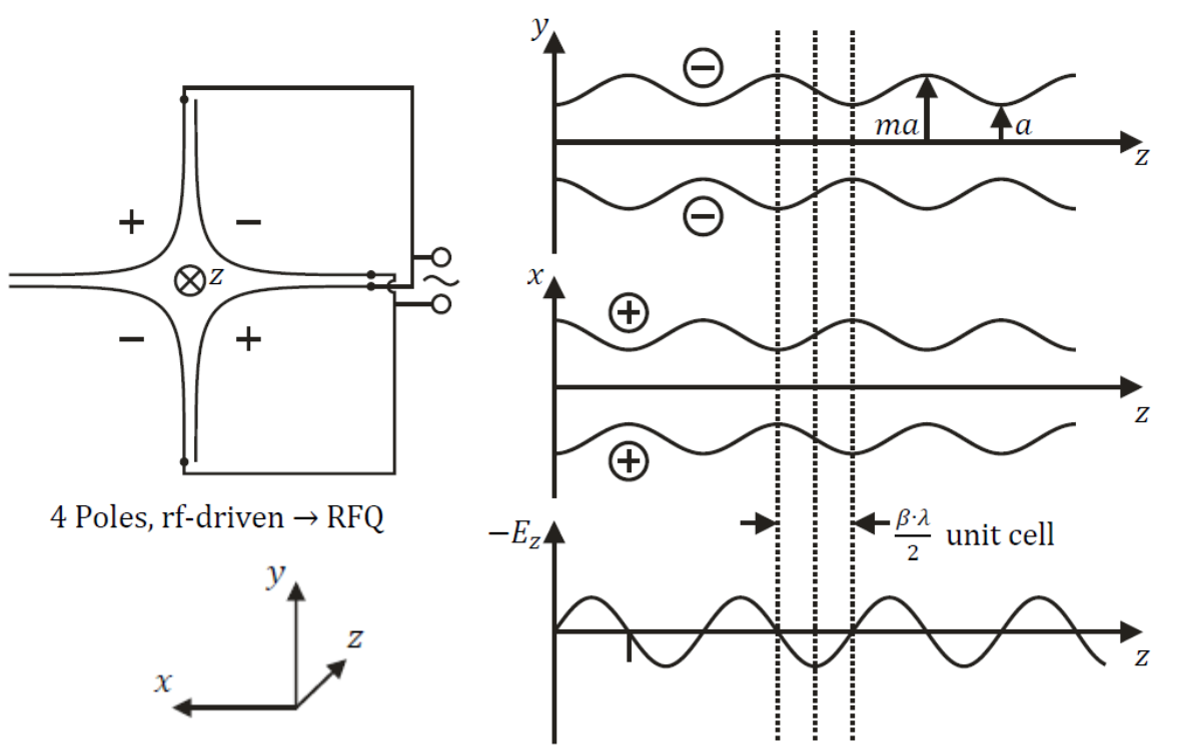}
	\caption{Scheme of electrodes inside of an radio-frequency-quadrupole.}
	\label{fig:RFQ}
\end{figure}

In the following, the work principle of an radio-frequency-quadrupole (RFQ) is summarized conceptually. A very good introduction can be found in~\cite{wangler}. The RFQ is a single cavity being installed behind an ion source. It simultaneously bunches, accelerates, and focuses an initial low energy dc-beam. Hence, it replaces a huge high dc-voltage device for pre-acceleration, several quadrupoles for focusing, and at least two bunchers at two different frequencies. Within the RFQ, the beam is guided by four electrodes (poles) as sketched in~Fig.~\ref{fig:RFQ}.
Opposing electrodes have electric potential of equal polarity and neighboring electrodes have potential of opposing polarity. The electrodes have hyperbolic shapes as yokes of magnetic quadrupoles. Being combined with a time wise sinusoidal potential imposed onto them, the propagating beam will be efficiently subject to a transverse FODO-lattice, which is overall focusing.

\begin{figure}
	\centering\includegraphics[width=.65\linewidth]{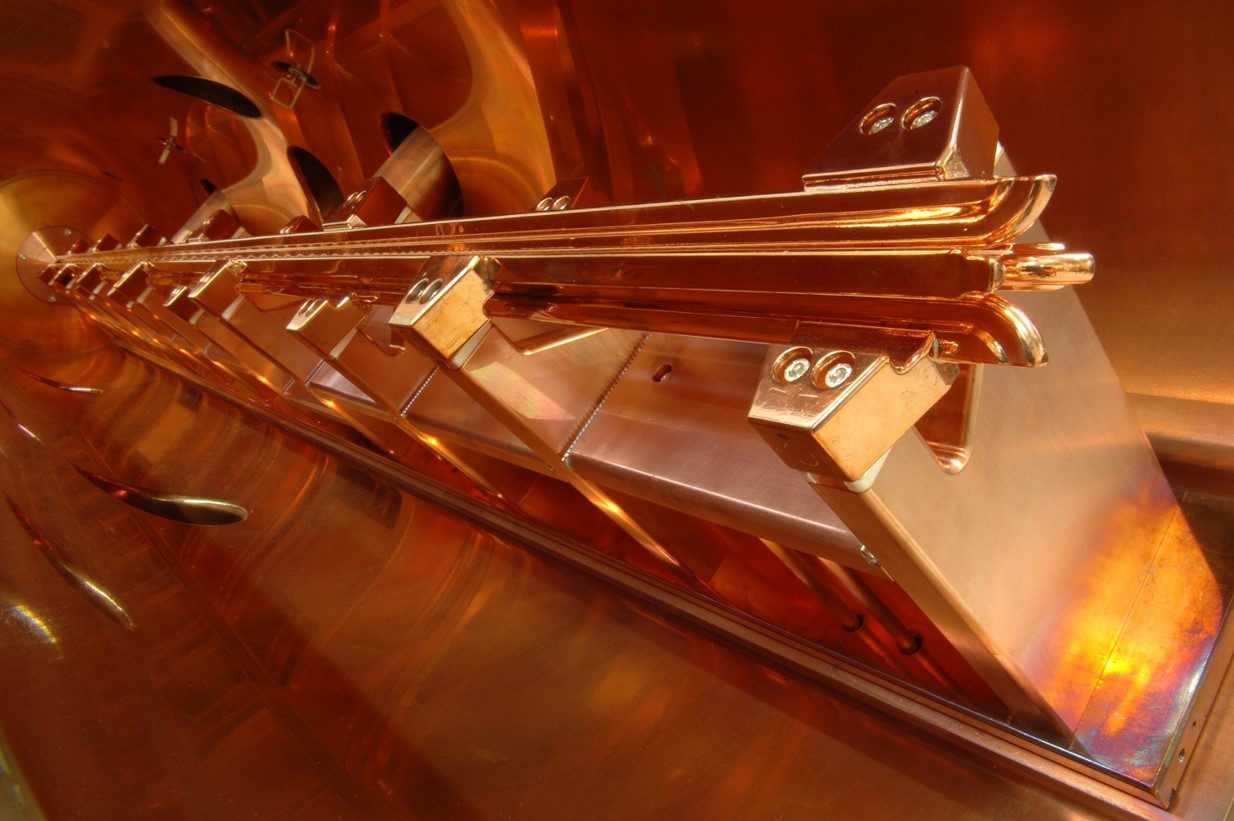}
	\caption{176~MHz radio-frequency-quadrupole of SARAF~\cite{saraf}. The electrode modulation increases smoothly from the entrance (right) to the exit~(left).}
	\label{fig:SARAF_RFQ}
\end{figure}

Additionally, the distances of the electrode pole tips from the beam axis are changed smoothly from an initial constant value to an oscillating value, i.e., the oscillation amplitude and wave length is increased along the axis. Opposing electrodes have the same distance to the axis, while the distance oscillation of the neighboring electrodes to the axis is shifted by usually~90° (in the figure this shift is~180°). This shift causes longitudinal electric field components along the axis, which together with the propagating beam and oscillation of the potential, leads to net longitudinal focusing. Accordingly, along the~RFQ the beam is smoothly transformed from a dc-beam to a chain of spatially separated bunches. The~smooth increase of the spatial wave length of the electrode modulation causes even permanent acceleration along the beam axis. Within an RFQ, the mentioned parameters need to be synchronized very well to each other. Once built, the only parameter being accessible to operators is the amplitude of the~potential (RFQ voltage). Figure~\ref{fig:SARAF_RFQ} shows the electrode modulation of the RFQ for the deuteron accelerator SARAF in Israel~\cite{saraf}. For further reading on RFQs, we refer to~\cite{wangler} as an excellent base to start from.

\section{Choice of cavity type}
Cavities and their Rf-power sources comprise a large fraction of up to 50\% of a linac project, i.e., several millions of Euros. Accordingly, the choice of cavity types should be made very carefully and be preceded by systematic analysis of its impact on many different criteria. Some of these may be "must haves" as pre-defined beam parameters at the linac exit, civil construction boundaries, and upper budget limits. Others are more soft and need to be weighted by the persons constructing, operating, and taking beam from the linac. Such criteria may be construction risks and duration, operational risks and flexibility, sensitivity to alignment errors and beam fluctuations, sizes and weights of components, requirements for spares and maintenance, on-site expertise, established collaboration partners, and many more.

For instance, at GSI the choice had to me made among four options. These have been evaluated w.r.t. the criteria mentioned above and tabled. Especially, the estimates on cost and construction efforts were done by the same persons in order to cancel out individual preferences for over/under-estimation. Final benchmarking of beam dynamics performance has been done by one single person using the same code for all options. In total, 34 criteria have been evaluated and each option has been summarized into a dedicated proposal. The material has been presented to an international expert committee, followed by final decision taking at the host lab. An analogue procedure is currently applied at~TRIUMF~\cite{abbaslou}.
\newpage
\section{Production issues}
Reasonable cavity radii should not exceed one meter to assure handling with reasonable efforts. The~cavity's interior dimensions shall be within a range that allows for access through humans and regular tools. This leads to the restriction of about
\begin{equation}
	\label{e_size}
	\text{10~cm}\,\leq\,\beta\lambda\,\leq\,\text{50~cm}\,.
\end{equation}
The choice of the frequency is practically ruled by the above equation combined with frequencies for which Rf-power sources are available on the market or which are used already at the laboratory. Picking a new frequency often is not required but it will impose considerable additional cost for re-developing something being basically at hand.

\begin{figure}[h!]
	\centering\includegraphics[width=.60\linewidth]{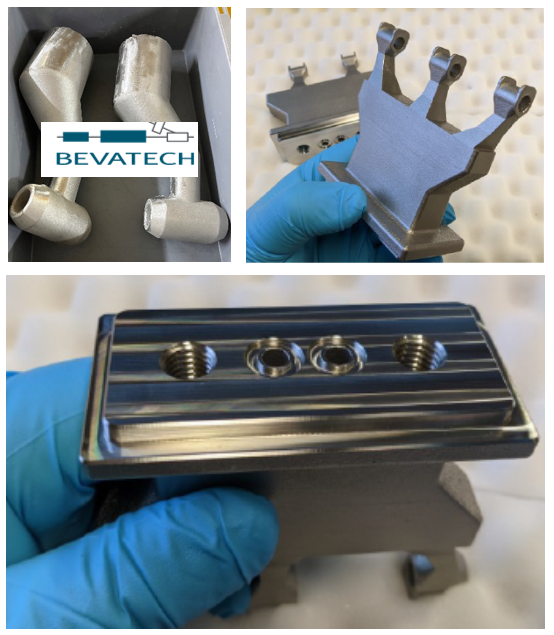}
	\caption{Examples for additive machining. Upper: Drift tubes for an IH-cavity~\cite{bevatech} and cooled girder with three stems~\cite{haehnel}; Lower: Cooled stems with drift tube for an IH-cavity buncher~\cite{haehnel}.}
	\label{fig:AddMan}
\end{figure}

It is emphasized that recent impressive progress made in additive machining (3d-printing) may considerably weaken the impact of the restrictions of~Eq.~(\ref{e_size}) w.r.t. the lower limit in size. Figure~\ref{fig:AddMan} for instance depicts stems and girders for IH-cavities being 3d-printed and including internal water cooling channels.

Another issue being addressed during the Rf-design is the maximum electric surface field strength. It must be sufficiently low to allow for long-term reliable operation without high voltage break downs, which even may destroy the cavity if occurring to frequently. In general, the maximum applicable field strength increases with the frequency. A semi-empirical scaling law from the 1940ies is the Kilpatrick-criterion for this strength
\begin{equation}
	\label{e_kilpatrick}
	f[\text{MHz}]\,=\,1.64\cdot E_K^2 e^{-\frac{8.5}{E_K}}\,,
\end{equation}
where $E_K$ is inserted in units of~MV/m. Modern designs allow for strengths in excess of this value by a~factor of up to~1.7 (J-PARC proton linac, 324~MHz) at frequencies above~100~MHz. At lower frequencies as 101~MHz, factors of up to six have been achieved in testing environments~(CERN lead injector). The actual choice depends on the required Rf-duty cycle as well as on the total amount of concerned surface. Last but not least it depends also on how conservative are the persons being in charge for the design. 

Once the theoretical Rf-design of cavities has been finished, the long way to the finally operating device is ahead. Occasionally, the according efforts are underestimated and eventual obstacles are considered as to be surpassed in one way or the other by appropriate engineering. This section tries to pick up some of the issues being encountered along that path.

\subsection{Material, tolerances, and alignment}
In order to minimize Ohmic surface current losses, inner cavity surface needs to be covered by a thin layer of gold, silver, or copper (mostly used). This imposes restrictions on the cavity raw material. In the past, mild steel has been the common choice. Compared to stainless it was considerably cheaper. Additionally, its heat conductivity is about twice larger w.r.t. stainless. However, nowadays the difference in cost is practically no more. In consequence, many high precision mechanical workshops waved mild steel and focus on stainless, since they must strictly separate the two materials. Cooling water channels within mild steel cause severe issues with erosion and clogging of the channels. Finally, many copper plating workshops focus on stainless as well in order not to spoil their basins from working with mild steel. However, some applications requiring high heat conductivity still use mild steel.

Proper definition of reasonable and required tolerances is an important step within the mechanical design. The natural relative bandwidth of Rf-power sources is about one per mill. For TM-mode cavities, the resonant frequency is dominated by the mean inner radius and by the radius' homogeneity (roundness). The relative error in radius corresponds to the relative error in resonant frequency and accordingly, the tolerance of the radius is within some few tens of a millimeter. Fabrication errors within the tolerances are compensated by plungers being moved inside of the cavity. They slightly reduce the~cavity volume and hence its effective radius. However, they cannot reduce the resonant frequency, and this must be included into the choice of the design radius. TM-cavities have the convenient property that the plungers' positions do practically not change the gap voltage distribution. In turn, drift tubes often house quadrupoles. The magnetic axis of the latter needs to be aligned onto the design beam axis with an accuracy of about~0.1~mm.

Instead, the resonant frequency of TE-mode cavities is dominated by the capacitance between the~drift tubes. Fine adjustment is done with plungers as for TM-cavities. However, the plungers' positions have considerable impact on the gap voltage distribution. Accordingly, the procedure of Rf-tuning might be quite lengthy especially for TE-cavities with many gaps. However, they have the advantage that there are no quadrupoles inside of the~drift tubes. Transverse alignment restrictions to drift tubes without quadrupoles are much more relaxed compared to those with quadrupoles.

\begin{figure}
	\centering\includegraphics[width=.70\linewidth]{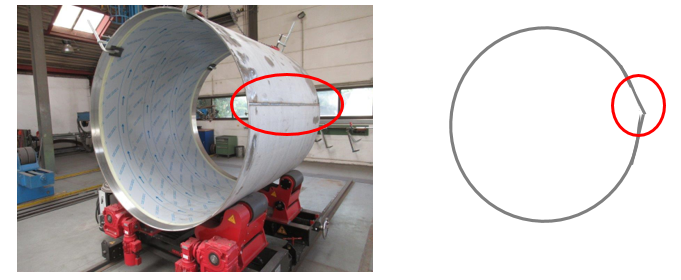}
	\caption{Rolling of a sheet of steel with a length of about six meters and thickness of 12~mm to a cavity mantle. At the weld the roundness is reduced.}
	\label{fig:rolling}
\end{figure}

Tight tolerances augment the production cost and reduce the amount of potential bidders. Critical tolerances are addressed already within the design phase through dedicated MWS simulations for instance. They may exclude some procedures during production or even allow for simple procedures not being considered previously. One example is the manufacturing of the outer cavity mantle. A simple way is to roll a flat sheet of steel as shown in~Fig.~\ref{fig:rolling}. However, it needs some welding tests to determine the~final mean radius and to minimize the un-avoidable pear-shape of the final mantle as well as to maximize the roundness. Alternatively, the inner mantle surface is (additionally) drilled. Drilling results into very precise mean radius and almost perfect roundness but also into considerable cost increase, especially for large cavities.

The cavity's final design dimensions should anticipate deformation from its intrinsic weight as well as from the outside pressure after evacuation. The latter concerns mainly the cavity end plates and hence the half drift tubes, which might by pushed towards the inside by some tens of a millimeter. Deformation from fluctuations of the cooling water temperature or the environment are to be avoided by proper temperature surveillance and control.

Alignment is critical for a quadrupole inside of a drift tube. The quadrupole's magnetic axis ($\vec{B}$=0) is measured relative to a fixed reference point at the tube's outer surface. Afterwards, the tube is installed into the cavity such that the magnetic axis is aligned to the design beam axis, even at the expense of an~eventually misaligned tube inside the cavity. Transverse magnetic fields impose much stronger restrictions to alignment w.r.t. longitudinal electric fields.

\subsection{Cooling}
For normal conducting cavities, the mean power to be cooled away is determined by the shunt impedance through
\begin{equation}
	P_{loss}\,\sim\,\frac{U_{cav}^2}{2R_s}\,.
\end{equation}
Cooling is achieved by integrating water channels into the cavity. A very useful rule-of-thumb states that a water flux of one liter per second, being imposed to a heat impact of 4.2~kW, is heated by one degree. Along the inner cavity surface, the power dissipation is not homogeneously distributed. It scales with the~local~$B^2$ and dedicated simulations can provide for according heating maps.

There are commercial codes to fully 3d-model the heat impact and the cooling channels. However, their use is very demanding, time consuming, and expensive since annual licenses can be up to~100~k€. For many practical applications, it turns out to be sufficient to apply simple analytical models~\cite{ramberger}. These just require the assumption of a constant temperature at the location of the cooling channels and the~basic equations of heat conduction, which need to be adapted to the specific geometry being modeled. The~following sub-sections sketch such modeling for the cavity mantle and for the end caps of a drift tube. The~detailed calculations can be found in~\cite{cooling}.

\subsubsection{Cavity mantle}
\begin{figure}
	\centering\includegraphics[width=.85\linewidth]{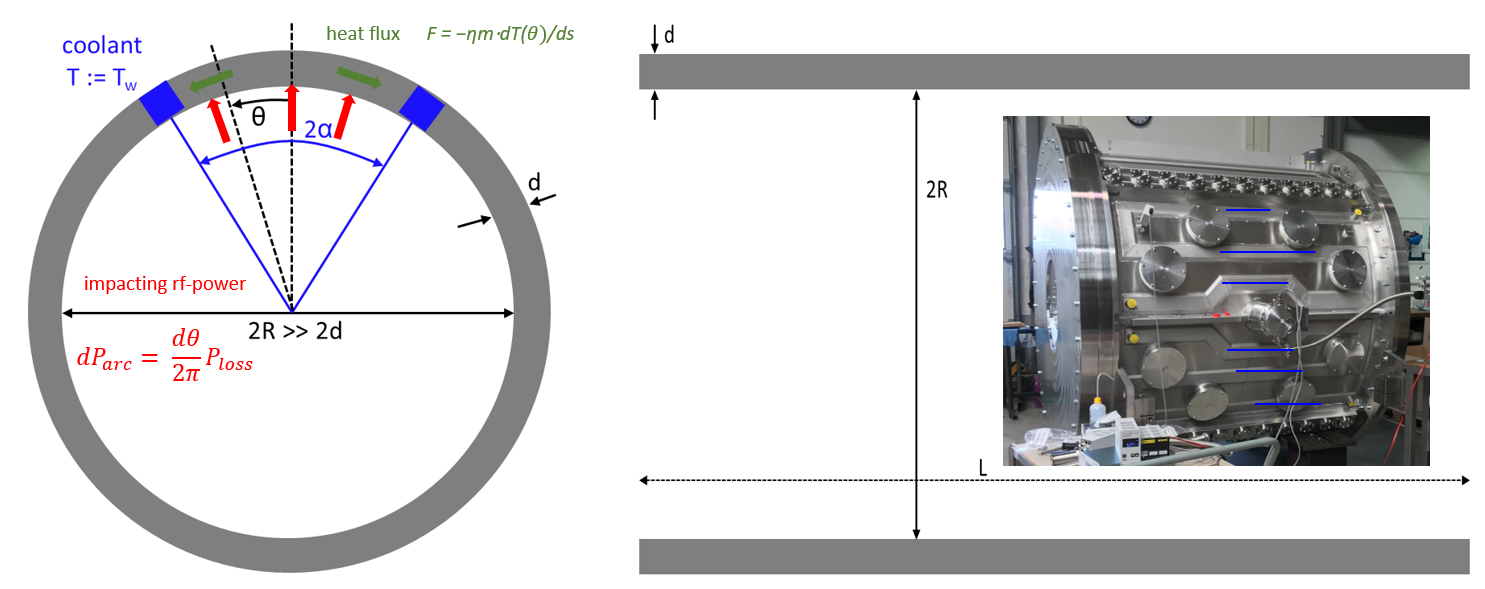}
	\caption{Analytic modeling of water cooling of a cavity mantle.}
	\label{fig:cooling_mantle}
\end{figure}

The geometric model for cooling the cavity mantel is drawn in Fig.~\ref{fig:cooling_mantle}.
Axial cooling channels are evenly distributed around the cavity mantle and they are separated by the angle~2$\alpha$. The power impacts from inside onto the mantle. Its thickness~$d$ is small compared to the radius~$R$. Inside of the mantle there is heat flux along the azimuth
\begin{equation}
	F\,=\,-\eta_md\cdot\frac{T(\theta)}{ds}
\end{equation}
towards the closest cooling channel, where $\eta_m$ is the mantle's heat conductivity and $T(\theta)$ is the local temperature. The dissipated power increment per arc increment~$d\theta$ is
\begin{equation}
	dP_{arc}\,=\,\frac{d\theta}{2\pi}P_{loss}\,,
\end{equation}
with the pessimistic assumption that the complete power loss is exclusively on the mantle. The local power flux is hence augmented by
\begin{equation}
	dF\,=\,dP_{arc}\frac{1}{d\cdot L}\,=\,\frac{P_{loss}d\theta}{2\pi d\cdot L}\,,
\end{equation}
with $L$ as length of the cavity along the beam axis. After integration, this is plugged into the equation of heat conduction delivering 
\begin{equation}
	T(\theta)\,=\,-\frac{P_{loss}R\theta ^2}{4\pi \eta _m d\cdot L}\,+\,T(\theta =0)\,.
\end{equation}
The maximum temperature on the mantle is defined as
\begin{equation}
	T_{max}\,:=\,T(\theta =0)
\end{equation}
and accordingly
\begin{equation}
	T_{max}\,=\,T_w\,+\,\frac{P_{loss}R\alpha ^2}{4\pi \eta _m d\cdot L}\,,
\end{equation}
with $T_w$ as the constant temperature of the cooling water. This equations allows to determine the required coolant distribution~$\alpha$ being required to keep the temperature below the tolerable value of~$T_{max}$.

\subsubsection{Drift tube end plate}
\begin{figure}
	\centering\includegraphics[width=.65\linewidth]{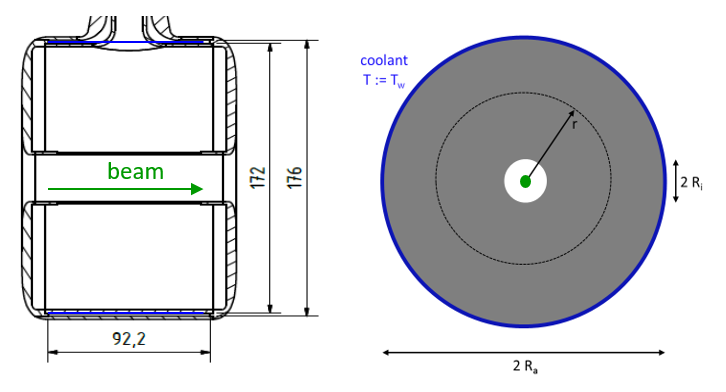}
	\caption{Analytic modeling of water cooling of a drift tube end cap.}
	\label{fig:cooling_DT}
\end{figure}

The scenario is sketched in Fig.~\ref{fig:cooling_DT}, approximating the end cap by a plane disk of thickness~$b$ with an~inner bore of radius~$R_i$ and an outer radius~$R_a$.
The mantle of the drift tube is cooled through a double-walled cylinder, while the end cap is just cooled indirectly, thus saving space and cost. However, it should be checked, whether there is no overheating. Firstly, the total amount of Rf-power~$P_c$ impacting on one tube end cap is estimated. This is done by scaling the mean total Rf-power loss on the full cavity~$P_{loss}$, as known from Rf-simulations, with the fraction of inner surface area taken by one end cap. Accordingly, the power loss on the end cap is estimated to
\begin{equation}
	\label{e_powscale}
	P_c\,=\,P_{loss}\cdot\frac{A_c}{A_{cav}}\,.
\end{equation}
This scaling is reasonable, as long as the power flux onto the cap is not significantly larger w.r.t. flux onto the cavity mantle, which takes the bulk of the total flux. Rf-simulations provide for such maps and hence allow to verify the assumption, or even to add an appropriate correction factor to the scaling.

Secondly, the temperature scaling~$T(r)$ on the cap is to be calculated. As previously,~$T(R_a):=T_w$ is assumed and the largest temperature is accordingly at~$r=R_i$. $T(r)$ is obtained through scaling laws from Maxwell's equation, namely starting from
\begin{equation}
	\vec{\nabla}\times\vec{B}\,\sim\,\dot{\vec{E}}\,,
\end{equation}
which being integrated over an area enclosed by~$r$ delivers
\begin{equation}
	\varoiint (\vec{\nabla}\times\vec{B})d\vec{A}\,\sim\,\dot{E}r^2
\end{equation}
and therefore $B\cdot 2\pi r\sim \dot{E}r^2$, or $B\sim r$.

One further uses $B\sim\dot{B}\sim I$, with~$I$ as Rf-induced surface current on the cap. Since the induced local surface power scales with~$I^2$, one finally obtains
\begin{equation}
	p\,\sim\,I^2\,\sim\,r^2\,,\text{i.e.},\, p(r)\,=\,a\cdot (r^2-R_i^2)
\end{equation}
taking into account the finite inner radius. Radial integration of this expression over the cap surface together with Eq.~(\ref{e_powscale}) allows for determination of the power scaling factor~$a$ in units of~[kW/m$^\text{4}$], i.e.,
\begin{equation}
	P_c\,=\,2\pi a\left[\frac{1}{4}R_a^4-\frac{1}{2}R_a^2 R_i^2\right]\,=\,P_{loss}\frac{A_c}{A_{cav}}\,.
\end{equation}
The incremental Rf-power imposed onto a ring of radius~$r$ and radial thickness of~$dr$ on the disk is given by
\begin{equation}
	dP\,=\,p(r)\cdot 2\pi r\,dr\,.
\end{equation}
Accordingly, at each radial position~$r$ the local power flux~$F$ towards the outer disk radius is augmented by
\begin{equation}
	dF\,=\,\frac{dP}{2\pi r\cdot b}\,=\,p(r)\frac{dr}{b}\,=\,a(r^2-R_i^2)\frac{dr}{b}\,,
\end{equation}
leading through integration to
\begin{equation}
	F(r)\,=\,\frac{a}{b}\left[\frac{1}{3}r^3-rR_i^2\right]\,+\,F_0\,.
\end{equation}
$F_0$ is given from~$F(R_i):=0$ and accordingly
\begin{equation}
    F(r)\,=\,\frac{a}{b}\left[\frac{1}{3}r^3-rR_i^2+\frac{2}{3}R_i^3\right]\,=\,-\eta _p\frac{dT(r)}{dr}\,,
\end{equation}
hence after integration delivering
\begin{equation}
	\label{e_Tr}
	T(r)\,=\,-\frac{a}{b\eta _c}\left[\frac{1}{12}r^4-\frac{1}{2}r^2R_i^2+\frac{2}{3}rR_i^3\right]\,+\,T_0\,.
\end{equation}
Using $T(R_a):=T_w$ one obtains
\begin{equation}
	T_0\,=\,T_w\,+\,\frac{a}{b\eta _c}\left[\frac{1}{12}R_a^4-\frac{1}{2}R_a^2R_i^2+\frac{2}{3}R_aR_i^3\right].
\end{equation}
After determination of $T_0$, the effective temperature increase $T(R_i)-T(R_a)$ follows from~Eq.~(\ref{e_Tr}).

Finally, it shall be mentioned that for a plated thin cap, the total heat conductivity is considerably augmented by its additional layer of copper of thickness~$d_{Cu}$, i.e.,
\begin{equation}
	\eta _c\,=\,\frac{\eta _{steel}\cdot b\,+\,\eta _{Cu}\cdot d_{Cu}}{b\,+\,d_{Cu}}
\end{equation}
resulting into enhanced cooling.

\section{Copper plating}
The specific Ohmic resistance of copper ($\rho\approx\,$0.017$\,\Omega$mm$^\text{2}$/m) is about a factor of 40 lower w.r.t. the~one of stainless steel. Accordingly, the amount of Rf-power loss is reduced through plating the inner cavity surface with copper. The thickness of this layer should be well above the~$1/e$-penetration depth of the~Rf-wave
\begin{equation}
	\delta\,=\,\sqrt{\frac{\rho}{\pi \mu _0 f}}\,,
\end{equation}
which is 21~$\mu$m at 10~MHz. In practice, layer thicknesses of about 15$\,\delta$ are applied in order to account for in-homogeneity of the layer thickness and hence to assure that also at the thinnest layer location, the~penetration depth is much smaller than the layer thickness. Alternatively, the whole cavity may be built from bulk copper. However, for large cavities this causes cost increase and mechanical deformation from intrinsic mass since copper is much softer than stainless.

Copper plating for accelerator application is a complex task. The number of experts around the world is very sparse. Many commercial plating companies can be found everywhere, but most of them focus on provision of nice-looking surfaces rather than on those being fit for the environment inside of an~operated Rf-cavity. The latter requires very clean copper, i.e., with resistance of less than~0.017$\,\Omega$mm$^\text{2}$/m. In order to minimize sparking, the final copper surface roughness~$\langle |h|\rangle $ should be below~1.0~$\mu$m, where $h$ is the local surface bump height w.r.t. the mean surface height. Additionally, strong adhesion is needed in connection with an outgassing rate of less than~10$^\text{-10}$~mbar~l~/~(cm$^\text{2}\,$s).

Prior to plating, the surface roughness should be less than 0.3~$\mu$m. The preferred raw material is stainless steel ("304L") and it must be free from voids. Plating itself is in several stages, i.e., pre-processing (some days), the actual plating (some hours), and post-processing (some days).
Pre-processing commences through removing of residual bumps from the surface and careful cleaning with appropriate chemicals. Afterwards, all surfaces of the cavity that must not be plated are masked. This applies especially to sealing surfaces. Holes, bores, and cooling channels must be closed such that no liquids from the plating baths can enter. This is followed by de-greasing of the remaining surfaces with appropriate chemicals. Surface activation is the last step of pre-processing.

\begin{figure}
	\centering\includegraphics[width=.35\linewidth]{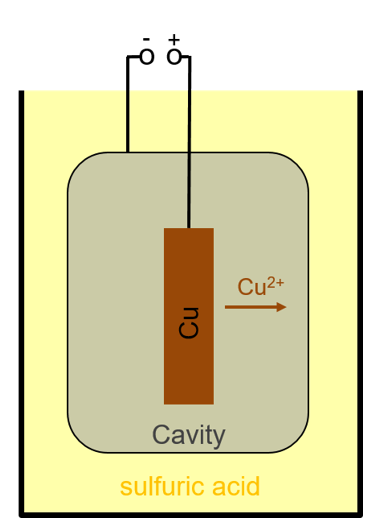}
	\caption{Scheme of galvanic copper plating of an inner cavity surface within a basin of sulfuric acid.}
	\label{fig:PlatingScheme}
\end{figure}

In case of stainless, prior to copper plating, all surfaces must be covered galvanically with a thin layer of nickel (some microns). Otherwise the copper does not stick. This step can be omitted with mild steel. Instead of nickel, gold may be used as well. However, dedicated measurements at 3.4~GHz did not reveal any difference w.r.t. the final conductivity~\cite{TE_report}. Afterwards, the actual copper plating is performed. Figure~\ref{fig:PlatingScheme} sketches the concept of galvanic plating.
The pre-processed cavity is fully immersed into a~chemical bath made from sulfuric acid and connected to negative potential, thus forming the cathode. The~anode comprises several electrodes from bulk copper. Definition, tailoring, and positioning of the single copper-electrodes within the cavity is one of the most demanding tasks of the whole procedure. The~electrodes' geometries and distribution determines the quality of the final layer and there are dedicated FEM codes to simulate the process. Improperly shaped or positioned anodes will cause blisters and/or parts of the surface to be over/under-plated. Once everything has been well-positioned, currents of up to one kAmp are applied. Ions of Cu$^\text{2+}$ drift from the copper electrodes to the cavity surface and hence provide for the aimed layer. It is not unusual to test the anodes' geometries through plating even several dummy cavities. Figure~\ref{fig:PlatingScheme} shows photographs of the plating process.
\begin{figure}
	\centering\includegraphics[width=.60\linewidth]{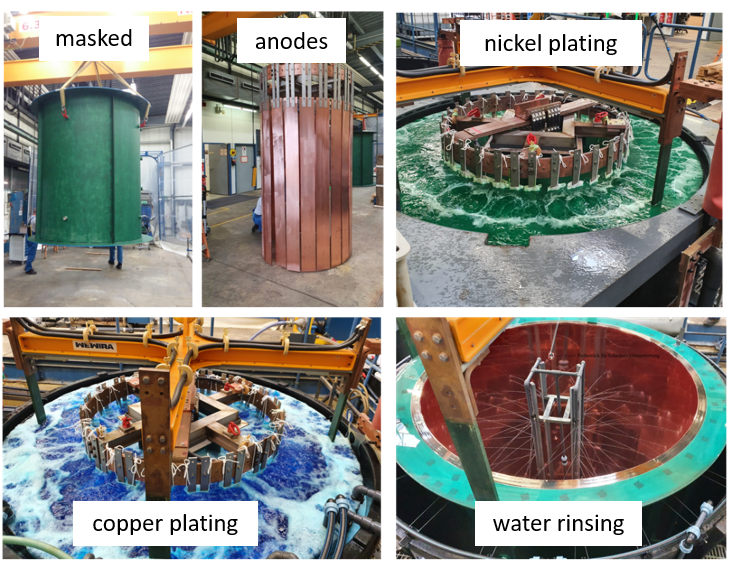}
	\caption{Galvanic plating of an inner dummy cavity mantle surface. From ul to lr: masked cavity dummy, copper anodes, nickel plating, copper plating, water rinsing.}
	\label{fig:PlatingBath}
\end{figure}

Post-processing comprises water-rinsing for cleaning, manual removal of blisters and bumps, and final manual polishing. Long-term storage until installation is done through applying pre-vacuum or sealing together with flooding by pure nitrogen.

\section{Rf-conditioning}
Albeit of careful polishing and cleaning of a new cavity, its surface is still covered with many tiny particles from dirt, dust, and mechanical spikes and grooves. These cause local surface field peaks, field emission, and thus melting of the perturbation. Drops from melting spread out locally and cause new (but smaller) imperfections. Evaporation leads to strong pressure increase, hence sparking, cavity de-tuning, Rf-break down, and finally to strong reflection of incoming Rf-power.

Initial perturbations are to be removed by proper Rf-conditioning, i.e., Rf-training of the cavity. Conditioning starts from operating at very low Rf-parameters as peak power level~$P$, pulse length~$\tau$, and repetition rate~$r$. The initially applied time-averaged Rf-power~$P\cdot\tau\cdot r$ is about five orders of magnitude lower compared to the aimed target parameter. Conditioning of a new cavity up to the design parameters may take some weeks to months. Cavities that once have been operated successfully at design parameters, after a considerable interruption of this operation scheme by some days, will need re-conditioning of some hours to days. Inter-mediate venting with air will prolong this period to days to weeks.

Rf-conditioning sort of resembles to cooking, since each expert has its individual recipe. However, there are some common features. The process is monitored by pressure and heat sensors and cooling water temperature measurements. Additionally, cameras, X-ray detectors, and residual gas spectrometers can be used. Rf-signals as coupled-in/out and reflected power are recorded and observed continuously. While observing these parameters, one of the Rf-parameters $P$, $\tau$, or~$r$ is increased smoothly. If the~pressure increases beyond about~10$^\text{-6}\,$mbar or break down of the reflected power occurs, the previously augmented parameter is lowered to reduce the pressure below about~10$^\text{-7}\,$mbar, to mitigate Rf-reflections, and to recover the previously achieved mean Rf-power inside of the cavity. Once the Rf-parameter has been conditioned to its design, the other parameters can be conditioned. However, switching among the~parameters can also be done at reasonable inter-mediate levels.
		
Cavities should be conditioned including margins of about 10\% beyond the design. The procedure usually closes with performing a long-term run of some days at these final levels. It is re-iterated that cavities become untrained if not being operated at these final parameters and according re-conditioning is required.

\begin{figure}
	\centering\includegraphics[width=.65\linewidth]{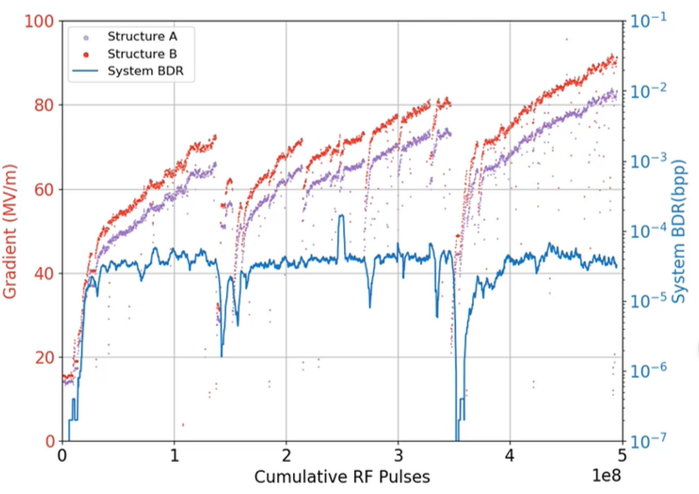}
	\caption{Example for achieved gradient (red and violet) and HV break down rate (blue) as functions of the number of applied Rf-pulses (figure taken from~\cite{millar}).}
	\label{fig:conditioning}
\end{figure}

Since long, several trials have been made to fully automate conditioning. Depending on the ambitions w.r.t. the aimed performance, these have been more or less successful. Within such procedures it has been realized, that surfaces are conditioned by those Rf-pulses that do not cause Rf-break down but rather during the subsequent recovery periods. A successful and well presented example can be found in the conference proceedings of L.~Millar et al.~\cite{millar}. Figure~\ref{fig:conditioning} plots the according long-term increase of the achieved cavity gradient as a function of the number of applied Rf-pulses.	


\newpage
\begin{thebibliography}{99}
\bibitem{TDR_Linac4}
Linac4 Technical Design Report, edited by F.~Gerigk and M.~Vretenar, CERN-AB-2006-084 ABP/RF, CERN, Geneva, Switzerland ~(2006).

\bibitem{ratzinger_1996}
U.~Ratzinger, The New GSI Pre-Stripper for High Current Heavy Ion Beams, Proc. of the XVIII Linear Accel. Conf., TU202, Geneva, Switzerland~(1996).

\bibitem{Clemente}
G.~Clemente, U.~Ratzinger, H.~Podlech, L.~Groening, R.~Brodhage, and W.~Barth, Development of room temperature crossbar-H-mode cavities for proton and ion acceleration in the low to medium beta range, Phys. Rev. Accel. \& Beams {\bf 14}, 110101~(2011).

\bibitem{wangler}
T.P.~Wangler, RF Linear Accelerators, 2nd edition, pp. 232, Wiley-VCH, Weinheim, Germany~(2008).

\bibitem{saraf}
D.~Berkovits et al., Operational Experience and Future Goals of the SARAF Proton / Deuteron Linac, Proc. of the XXVI Linear Accel. Conf., MO1A01, Tel Aviv, Israel~(2012).

\bibitem{abbaslou}
M.~Abbaslou and B.~Laxdal, Design Considerations for a Proton Linac for a Compact Accelerator Based Neutron Source, Proc. of the XXII Linear Accel. Conf., FR1AA05, Liverpool, UK~(2022).

\bibitem{bevatech}
https://bevatech.com/

\bibitem{haehnel}
H.~H\"ahnel, A.~Ate\c{s}, U.~Ratzinger, A 3D Printed IH-Type Linac Structure, Proc. of the HIAT Conf., TU1C4, Darmstadt, Germany~(2022).

\bibitem{ramberger}
S.~Ramberger, priv. communication.

\bibitem{cooling}
L.~Groening, Analytic estimate of cooling of a cavity mantle and drift tube end plates, ALV2\_note\_20190218, GSI~(2019).

\bibitem{TE_report}
CERN TE Technology Department, Copper Plated Discs on Nickel and Gold for GSI, CERN report~(2023).

\bibitem{millar}
L.~Millar at al., High Power Conditioning and Breakdown Studies in Coupled Accelerating Structures, Proc. of the 30th Linear Accel. Conf., TUPAB076, Liverpool, UK~(2020).


\end{thebibliography}
\end{document}